\documentclass{knawproc}

\usepackage{epsfig}

\begin{document}

% The ``opening'' environment takes care of title, author and headlines
\begin{opening}

\title{Galaxies and clusters 
around and in front of radio sources at $0.5<z<4.5$}

\author{Mark Lacy}

\runningtitle{Galaxies, clusters and radio sources}
\runningauthor{M.\ Lacy}

\end{opening}

\begin{center}
Astrophysics, Dept.\ of Physics, Keble Road, Oxford
\end{center}

%%%%%%%%%%%%%%%%%%%%%%%%%%%%%%%%%%%%%%%%%%%%%%%%%%%%%%%%%%%%%%%%%%%%%%%%%%%%%%%

\begin{abstract}
Some recent work on radio source fields at $z\sim 0.7$ and $z\sim 4$ is 
discussed. At $z\sim 0.7$ we find that radio-loud quasars are typically
found in moderately rich environments independent of radio luminosity, 
consistent with previous results at $z\sim 0.5$. In the field of the 
$z=4.41$ radio galaxy 6C0140+326 we find several candidate $z>4$ 
galaxies using the Lyman-break technique, two of which have detectable
UV absorption features. In two $z>4$ radio galaxies,
we find evidence for gravitational lensing affecting the fluxes by a few
tens of percent, although we cannot rule out unusual 
lensing events which have larger magnifications associated with
them. A simple calculation suggests lensing of $z>4$ radio sources
should be very common. 
\end{abstract}

%%%%%%%%%%%%%%%%%%%%%%%%%%%%%%%%%%%%%%%%%%%%%%%%%%%%%%%%%%%%%%%%%%%%%%%%%%%%%%%

\section{Introduction}

The discoveries that powerful FRII radio galaxies and radio-loud quasars are 
frequently found in moderately rich clusters by $z\sim 0.5$, in contrast
to the situation at low redshifts, represent a large evolutionary change
in the relatively recent past (Hill \& Lilly 1991; Ellingson, Yee
\& Green 1991). Studying the cluster environments at higher redshifts 
is clearly important, and is possible in the optical up to $z\sim 0.8$.

At $z\stackrel{>}{_{\sim}}0.8$ when the 4000A break redshifts out of the 
optical window, cluster galaxies become hard to 
distinguish from the background.
This can be partly improved by observing in the near-infrared, but even 
here the slow change of angular size distance with redshift means that
clusters of a given physical size become no more compact on the sky. At the
highest redshifts, spectral methods of identifying companion galaxies become 
necessary, for example narrow-band or, at sufficiently high 
redshift, Lyman-limit techniques. 

Foreground clusters may be important too. Claims of
statistical correlations
between luminous radio sources and foreground galaxies (Benitez et al.\ 
1997; Hammer \& LeFevre 1990) are supported by correlations
seen for different classes of AGN with foreground galaxy and cluster catalogues
(e.g.\ Bartsch, Barthelmann \& Schneider 1997; Rodrigues-Williams \& Hogan
1994) and by the detection of shear
fields around a few $z\sim 1$ radio sources (Fort et al.\ 1996;
Schneider et al.\ 1997). The nature of the lensing population,
and how it is able to produce these effects whilst retaining a plausible 
form for the AGN luminosity function remains a mystery, but large (i.e.\ 
cluster-sized) lenses could lens both radio sources, any close companions, and
any even higher redshift objects in the field.

In this paper I begin by describing some early results of optical 
observations of the cluster environments of a sample of radio-loud quasars 
at $z\sim 0.7$ made with the NOT, as part of an on-going programme to 
study the cluster environments of radio sources and radio-quiet quasars
at $z\sim 0.7$, in collaboration with Margrethe Wold and Per Lilje.
I then discuss some work currently under way with my student Robin
Stevens on $z>4$ Lyman limit systems, before discussing weak lensing results
on the two most distant radio galaxies.

\section{Clusters at $z\sim 0.7$}

We have embarked on a programme of imaging the fields of $z\sim 0.7$ 
AGN using the Nordic Optical Telescope
to measure the richnesses of their environments. We have taken data
on the fields of radio-loud quasars, radio-quiet quasars and radio galaxies, 
selected so as to span as large a range in AGN luminosity as possible ($>10$
in all cases). The data on the radio-loud quasars has been fully analysed
(Wold 1996) and reveals that the environments of $z\sim 0.7$ radio-loud
quasars are similar to those of the $z\sim 0.5$ quasars studied by Ellingson 
et al.\ (1991), i.e.\ these objects are frequently found in moderately 
rich clusters. There is no strong dependence on the environment with radio
luminosity over a range of $\sim 100$ in radio power, consistent 
with the results for $z\sim 0.5$ radio galaxies found by 
Hill \& Lilly (1991). Our results will be 
discussed in more detail in Wold et al.\ (1998, in prep).

These results are an interesting contrast to those of Roche et al.\ (1997)
for 6C radio galaxies. In these, which are only about four times fainter than 
3C radio galaxies at the same redshift, they find that the 
scale-sizes and luminosities are significantly smaller than those measured 
for 3C radio galaxies by Best et al.\ (1997). This would
suggest that 3C galaxies are more massive than 6C ones and are therefore 
likely to reside in richer environments. This implies a
significant change in the dependence of 
clustering on radio luminosity over a relatively short period of cosmic time.

\section{Lyman break galaxies at $z>4$}

With my PhD student in Oxford, Robin Stevens, I have been working on detecting
Lyman-break galaxies in the fields of $z=3.4-4.4$ radio sources. We have 
imaged the fields of B2 0902+34 ($z=3.4$), 4C41.17 ($z=3.8$), 8C1435+635
($z=4.25$) and 6C0140+326 ($z=4.41$) with the WHT prime focus (field area
$\approx 40$arcmin$^2$). We have used $U$-band as the short
wavelength filter for the two $3.4<z<4$ sources and $B$-band for the 
others. Spectroscopy with LDSS2 on the WHT of the 6C0140+326 field has
revealed eight  objects with low-dispersion spectra consistent with them
being $z>4$ galaxies, two of which have several plausible absorption features
placing them at a redshift of $4.02$ and $4.16$ respectively
(see Spinrad, these proceedings, for a discussion of UV absorption features
seen in high-$z$ starbursts).
These observations are discussed in more detail in 
Stevens, Lacy \& Rawlings (1997 and in prep).

\section{Lensing of high redshift radio galaxies}

We have recently investigated the possibility that the 
two most distant radio galaxies (to which the cross-section to lensing should 
be highest) may be gravitationally amplified. The questions we 
wished to address were: could the magnification be high enough to significantly
affect estimates of source properties (e.g.\ stellar content, star
formation rate and morphology), 
and could it distort the radio source luminosity function
sufficiently to boost the numbers of luminous radio sources at high-$z$?

\subsection{8C1435+635}

This $z=4.25$ radio galaxy has a $z=0.24$ galaxy 
projected just beyond the south east radio
lobe. This is close to a larger, $\approx L_*$ galaxy at $z=0.23$, 
possibly in the same group or cluster (Lacy et al., 1994). 
Another group of three galaxies was found 
at $z\approx 0.35$. (All these were found in long-slit spectra designed to 
measure the radio source redshift, and as yet no systematic redshift survey 
of the field has been made.) With a deep $I$-band image from the WHT, we 
made a weak lensing study of this field, using the technique of 
Seitz \& Schneider (1995). A marginally 
significant (3.5-$\sigma$) peak is present in the recovered projected mass 
distribution, offset by about 50 arcsec from the radio galaxy (Fig.\ 1).
Apart from the groups already mentioned there seems to be no obvious 
optical cluster in the field, and  
no X-ray emission is seen in this field in a deep ROSAT exposure.
This suggests either a high redshift for the cluster providing the bulk of
the weak lensing signal, or a ``dark cluster'' with a high mass:light ratio. 

\begin{figure}[h]
\includegraphics[scale=0.6]{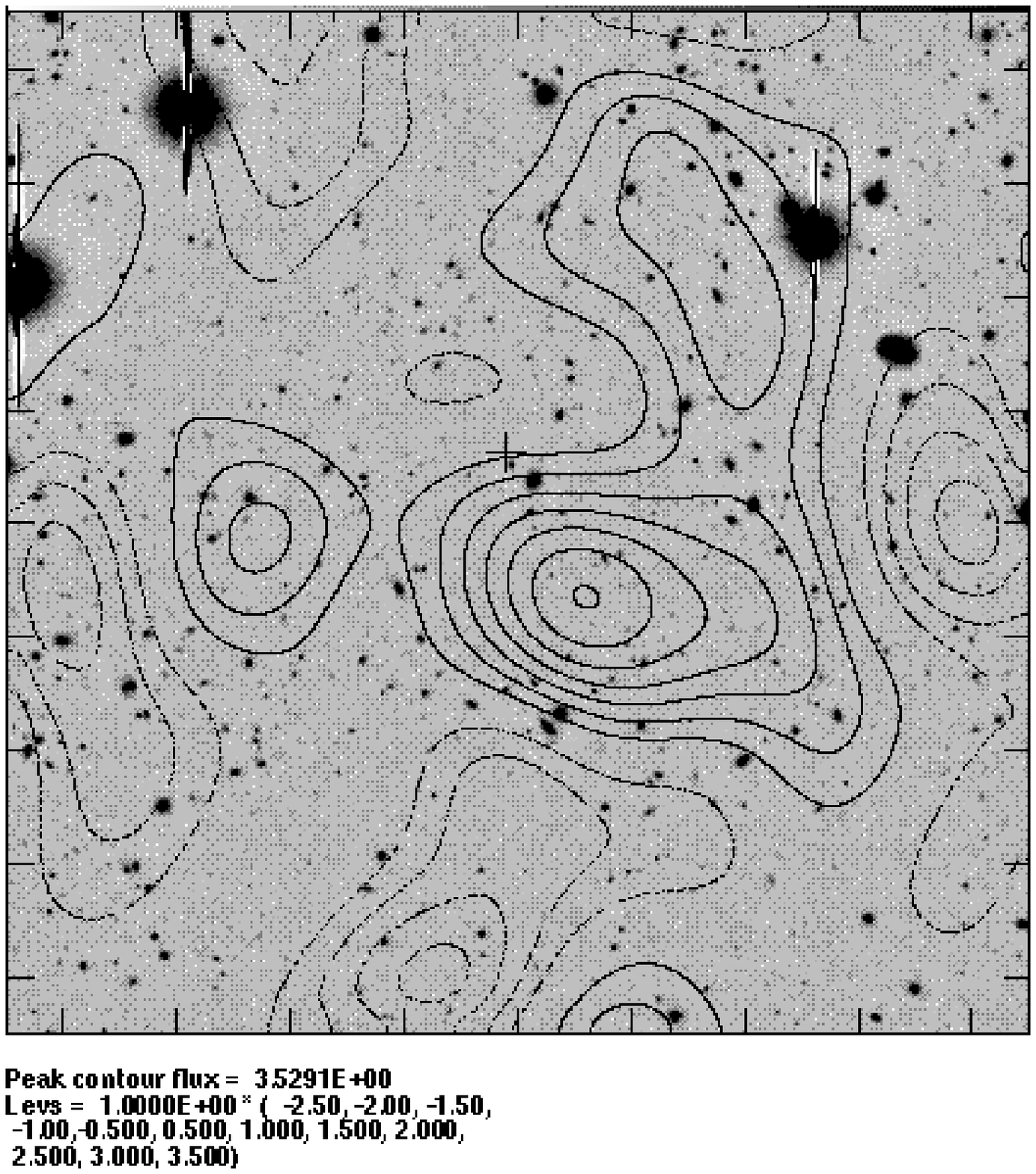}
\caption{The greyscale of our WHT $I$-band image of 8C1435+635 with 
significance contours of the projected mass distribution superposed,
spaced at 0.5$\sigma$ intervals (negative contours shown dashed). The
field size is 6 arcmin square, and the
approximate position of the radio source is marked with a cross}
\end{figure}

\subsection{6C0140+326}

The evidence for lensing in this source is more secure. Again, a nearby 
(in projection) galaxy was noted in the discovery paper (Rawlings et al.\
1996) at $z=0.93$. This was sufficient to produce significant 
lensing on its own, revealed in our 5GHz MERLIN image (Fig.\ 2), where
both the hotspots are elongated in the direction expected if they
are being lensed by the $z=0.93$ galaxy. Although the exact magnification 
factor is hard to estimate due to the fact that the source structure is
unknown, a simple singular isothermal sphere (SIS) model 
can be used to obtain a rough magnification if 
initially circular hotspots are assumed. The maximal
magnification from an SIS can be found by placing an 
isothermal sphere with an Einstein radius of 0.9 arcsec at the nominal
lens position (uncertain by $\approx \pm 0.3$ arcsec). This is 
consistent with the galaxy 
luminosity and at our best estimate distance from the source, and assuming
initially circular hotspots produces magnifications of 2 for the
eastern hotspot and 1.4 for the west (overall 1.7) (see Fig.\ 2). 
Models with elliptical isothermal potentials
aligned in the direction of the observed lensing galaxy orientation 
give higher overall magnifications of $\approx 2$. 

\begin{figure}[ht]
\includegraphics{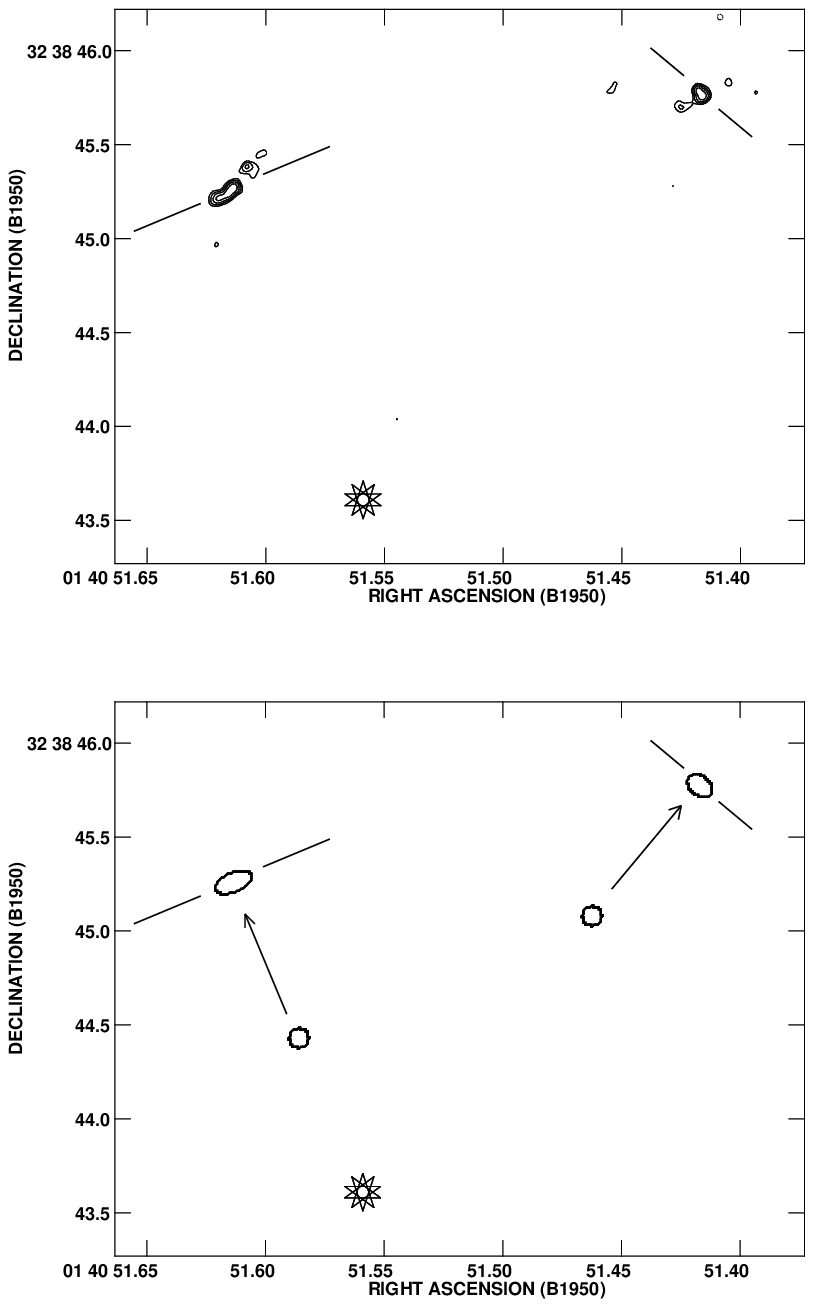}
\caption{Top: MERLIN image of the $z=4.41$ radio source 6C0140+326, made at 
5GHz with 50maS resolution, contours at factors of $\sqrt{2}$ from
0.3mJy. Below: a model in which two initially circular hotspots are
lensed into arcs by an SIS with Einstein
radius 0.9 arcsec centered on the position of the lens (indicated by a
star in the top and bottom figures). Tangent lines are drawn either
side of the two model hotspots, and are repeated in the upper panel to
guide the eye.}
\end{figure}

\subsection{Is lensing important?}

Some of the best constraints on lensing of AGN come from studies in the radio: 
with typical lens mass profiles
(SISs or the CDM profile of Navarro, Frenk \& White 1997) we would expect
large numbers of multiple images or radio lobes imaged into rings if
overall source magnifications frequently exceed $A \sim 2$ [although
Kochanek \& Lawrence (1990) suggest a large fraction of smaller radio rings
are missed due to resolution effects]. 

An important caveat to this though is that additional sources of magnification 
which have scale sizes much larger than the size of the radio source 
could in principle produce significant magnification without image splitting. 
The effects of a general lensing mass distribution 
can be broken down into two components, a
shear component due to matter distributed inhomogeneously outside the beam 
with which you observe 
a distant object (thus a point-mass lens outside the beam 
is a pure shear lens), and 
a convergence term due to matter within the beam (thus a uniform mass
sheet is a pure convergence lens). Most realistic lensing mass distributions, 
such as the SIS model described above have contributions 
of both shear and convergence to the total amplification, but it is possible
to conceive of other mass distributions in which either the shear or the 
convergence term dominates.

Large-scale shear (i.e.\ shear with a scale size comparable to that of the 
radio source) should be recognisable in terms of further distortion 
of the image, and the good match of the SIS model to the 
image of 6C0140+326 suggests that it must be fairly small in this
case\footnote{Large-scale shear would be necessary for the alignment effect
to be produced via lensing (e.g.\
Le F\`{e}vre et al.\ 1988). Although it is possible
that in some cases such as 6C0140+326 the alignment could be enhanced
by lensing it is highly unlikely that many $z\sim 1$ radio sources could
be sufficiently distorted by large-scale shear to reproduce the 
alignment effect without many cases of multiple images being seen, 
even if the lenses themselves have a high mass:light ratio.}. Much
harder to deal with is the possibility of a mass sheet associated with
a cluster. This could magnify an image without producing
distortion. Williams \& Lewis (1997) have 
recently discussed the possibility of highly magnified but unsheared
single images which could be formed in the centres of clusters. These depend
on the dark-matter profiles having significant core radii (and therefore
effectively acting as mass sheets). 
Weak lensing studies of a large enough field should 
eventually reach regions where the shear from any lensing cluster is 
significant though, and we could therefore infer the presence of a lensing 
cluster from this, as we may indeed have done in the case of 
8C1435+635.

If, however, we proceed on the assumption that what we are seeing is indeed
only fairly small magnifications, the next question is the frequency with 
which we expect to see this effect in flux-limited samples of radio sources.
The frequency of lensing in a flux-limited sample may be estimated using
a model for the lensing population combined with an estimate of the 
magnification distribution produced by this population and 
the radio source luminosity function. We will follow Peacock (1982) in 
assuming a population of singular isothermal lenses. For this
population he derives
an analytic approximation to the magnification distribution $F(A)$ which 
has a tail $\propto A^{-3}$, and turns over near $A=1$ according to 
a lognormal distribution with a scatter $A^*-1$ which represents the 
spread in amplifications about unity due to multiple lensing events. 
Four conditions on this function allow it to be fairly well constrained;
$F(A)=0$ at $A=0$; $\int_0^{\infty} F(A) {\rm d}A =1$ (normalisation); 
$\int_0^{\infty} AF(A) {\rm d}A =1$ (mean amplification must be unity), 
and finally the strong lensing tail should give the same number of 
multiple images of point sources 
as predicted by models of strong lensing statistics 
(e.g.\ Kochanek \& Blandford 1987). The other important parameter of this 
model is the maximum magnification due to lensing, which is 
proportional to the ratio of the source size to the Einstein ring diameter,
and is 
\[ A_{\rm max} = 70 (D_{\rm s}/(c/H_0))(d/{\rm kpc})^{-1}\]
for our model, where $D_{\rm s}$ is the source angular size distance and 
$d$ is the source size (Peacock 1982).

Combining this with a power-law luminosity function of the form 
$\phi(L) \propto L^{-\beta}$, we can  
estimate the factor $b$ by which lensed sources are overrepresented in a
flux-limited sample:
\[ b = \int_0^{\infty} A^{\beta-1} F(A) {\rm d}A \]
(Peacock 1982). To estimate the mean magnification of a lensed source
in a flux-limited sample we need to evaluate
\[<A> = \frac{1}{b}\int_0^{\infty} A^{\beta} F(A) {\rm d}A \]

We also need to allow for the fact that
cross-section to lensing of radio sources is significantly higher
than that of optical quasars, because the sizes of the regions of highest 
brightness radio emission are 
often significant compared to the Einstein radius of the 
lens (Kochanek \& Lawrence 1990). This increases the 
cross-section to lensing in the strong lensing tail by a factor
\[ f_{\rm HS} =  
\left(1+g_1\frac{l_1 + l_2}{r_0} + g_2 \frac{l_1 l_2}{r_0^2} \right) \]
where $l_1$ and $l_2$ are the major and minor axis lengths of the source,
$g_1$ and $g_2$ are constants equal to 5/2 and 10 respectively in
a spatially flat Universe and $r_0 = 1.4 \sigma_{220}^2$ arcsec where
$\sigma_{220}$ is the velocity dispersion in units of 220 kms$^{-1}$
[Kochanek \& Lawrence (1990); Kochanek 1993]. Note that the
definition of source size here is relative to the Einstein radius
$\theta_{\rm E}$ of
the lens. Hence, in the case of FRII sources,  
if the hotspot separation $\theta_{\rm R} >>
\theta_{\rm E}$, lensing events on the two
hotspots can be considered independently. 

To estimate the importance of lensing on the most distant radio sources, we
can take 6C0140+326 as a specific example. At $z=4.4$ the cross-section 
to multiple imaging of a point source in an $\Omega=1$
cosmology is $\approx 2.6\times 10^{-3}$ (Kochanek \& Lawrence 1990). 
This source is an example where
$\theta_{\rm R}\sim \theta_{\rm E}$, and the effective 
source size is debatable. We have decided to consider the two
hotspots independently for the purposes of this analysis. Another complication
is that the hotspot size used should
strictly speaking be the size at the selection frequency, 151 MHz in
the case of 6C0140+326. We will make the (probably
pessimistic) assumption that the hotspot sizes are no larger than seen 
in high frequency radio maps, $\approx 0.1$ arcsec in diameter for
6C0140+326, giving $f_{\rm HS} \approx 1.2$. 
We then double the cross-section to allow for both
hotspots reaching $\approx 6.2\times 10^{-3}$. 
Our hotspot size 
gives $A_{\rm max}\approx 24$. Using these values we obtain 
$b=1.09$ and $<A>=2.0$ for $\beta=3$. Thus the luminosity function is 
little changed by the effect of lensing, primarily due to the flux
conservation condition, but the mean magnification can be comparatively high, 
basically because the distribution of luminosity-function-weighted
amplifications is very flat in the high-$A$ tail, with the areas of the peak
about unity and the tail out to $A_{\rm max}$ 
comparable. If we take these numbers as typical, we might expect to see
about half the $z\approx 4$ sources in a flux-limited sample
significantly lensed, with a wide range in magnification factors.

The degree of lensing is very sensitive to the 
slope of the luminosity function. In Table 1 we list the enhancement factors
and $<A>$ as a function of luminosity function slope. 
It can be seen that $<A>$ increases with $\beta$ too though. 
In fact the lack of obviously
highly-magnified images suggests $\beta$ is probably $\leq 3$ at $z\sim 4$ 
to keep the mean magnification in the strong lensing tail relatively low.

Including clusters in the analysis
will add further to the lensing structures available. A hint that clusters 
may be important comes from our study of 8C1435+635, and Peacock (1982) 
estimates that clusters can provide as much as a factor of two in the 
optical depth. Also, cosmology is potentially 
very important. A low cosmic density or a significant cosmological 
constant both increases the cross-section to lensing to point masses and,
in the case of non-flat cosmologies, the 
factor $f_{\rm HS}$ by which extended sources have larger cross-sections
(Kochanek 1993). 

\begin{table}
\caption{Lensing enhancement factors and mean magnifications as a function
of $\beta$}
\begin{tabular}{lrrrr}
$\beta$ & 2.5  & 3.0 &3.5 &4.0 \\ \hline
$b$     & 1.02 & 1.09&1.31&2.18\\
$<A>$   & 1.29 & 2.0 &4.87&14.2\\
\end{tabular}
\end{table}

Careful studies of both weak and strong lensing of complete samples 
of distant radio sources should eventually 
allow studies of lens statistics with a much larger sample of 
objects than available for conventional lensing studies, besides
potentially being able to find new classes of lensing objects. It may
also prove possible to use lensing arguments 
to constrain the luminosity function
at redshifts where too few sources are known to measure it directly. 

\begin{acknow}
I thank John Peacock for a helpful discussion which prompted me to
attempt a more in-depth analysis of the effects of lensing, and
Lindsay King and
Susan Ridgway for both helpful discussions and reading the manuscript.
\end{acknow}

\end{document}